\documentstyle[twoside,fleqn,espcrc2]{article}

\newcommand{\AmS}{{\protect\the\textfont2
 A\kern-.1667em\lower.5ex\hbox{M}\kern-.125emS}}
\title{Doping Dependence of the Magnetic Resonance Peak in 
$\rm YBa_2 Cu_3 O_{6+x}$}
\author{B. Keimer\address{Dept. of Physics, Princeton 
University, Princeton, NJ 08544, USA}\thanks{Supported by NSF-DMR94-00362.}
, H.F. Fong$\rm ^a$, S.H. Lee\address{Reactor Division, 
NIST, Gaithersburg, MD 20899, USA}, D.L. Milius$\rm ^c$, 
I.A. Aksay\address{Dept. of Chemical Engineering, Princeton University, 
Princeton, NJ 08544, USA}}
      
\begin{document}
\begin{abstract}
We report inelastic neutron scattering experiments on the doping 
dependence of the energy and spectral weight of the sharp magnetic 
resonance peak in $\rm YBa_2 Cu_3 O_{6+x}$. These measurements 
also shed light on the relationship between the magnetic excitations 
in the normal and superconducting states.
\end{abstract}
\maketitle

Magnetic excitations in high temperature superconductors have 
been intensively studied experimentally and theoretically for a 
number of years as they provide a direct and incisive probe of 
correlation effects in the cuprates. These efforts have been 
redoubled after the discovery of a sharp magnetic collective mode 
in $\rm YBa_2 Cu_3 O_7$ by inelastic neutron scattering 
\cite{fong96,bourges96,mook93}. This 
mode is strongly coupled to superconductivity in this material; in 
fact, it is only present in the superconducting state and disappears 
in the normal state \cite{fong96}. Two different mechanisms, with various 
modifications, have been proposed to explain this observation. First, 
it may be a consequence of the pileup of electronic states above the 
superconducting energy gap which compensates for the loss of 
states below the gap. Both a $d$-wave BCS gap function with strong 
Coulomb correlations \cite{levin95,millis96,scalapino94} and the (non-BCS) 
gap function resulting from the interlayer pair tunneling model of 
superconductivity \cite{chakravarty97} can account for the sharpness 
of the mode in both wavevector {\bf q} 
and energy $\hbar \omega$. Second, superconductivity may 
provide a matrix element (through particle-hole mixing) that 
couples a preexisting collective mode to the external probe, 
magnetic neutron scattering \cite{zhang95}. 
 
Further experimental information is clearly necessary in order to 
distinguish between these fundamentally different interpretations.
Since the doping dependence of the superconducting energy gap has 
recently been determined independently by angle-resolved 
photoemission \cite{harris96}, the doping dependence of the collective mode may 
provide additional insights into the mechanism reponsible for 
coupling of the spin excitations to superconductivity. Further, the 
normal-state spin susceptibility in $\rm YBa_2 Cu_3 O_7$ is too 
small to be reliably determined by present neutron scattering 
techniques. In underdoped $\rm YBa_2 Cu_3 O_{6+x}$, however, a 
sizeable normal-state susceptibility has been observed in previous 
experiments, so that the relationship between the sharp resonance 
resonance mode in the superconducting state and the normal-state 
excitation spectrum can be probed.
 
The experiments were conducted on the H4M, H7 and H8 thermal 
triple-axis spectrometers at the High Flux Beam Reactor at 
Brookhaven and at the SPINS cold-neutron triple axis spectrometer 
at NIST. At Brookhaven we used neutrons with 30.5 meV final energy 
and adjusted the energy resolution to $\sim7$ meV by collimating 
the beam. At NIST we used neutrons with 3.5 meV final energy, with 
typically $\sim 0.1$ meV energy resolution. In order to convert the 
observed intensity to the dynamical susceptibility $\chi''({\bf 
q},\omega)$ in absolute units, the magnetic cross section was 
calibrated to transverse acoustic phonons around the (004) nuclear 
Bragg reflection at low energies and to an oxygen vibration of 
energy 42.5 meV at high energies.

\begin{figure}[htb]
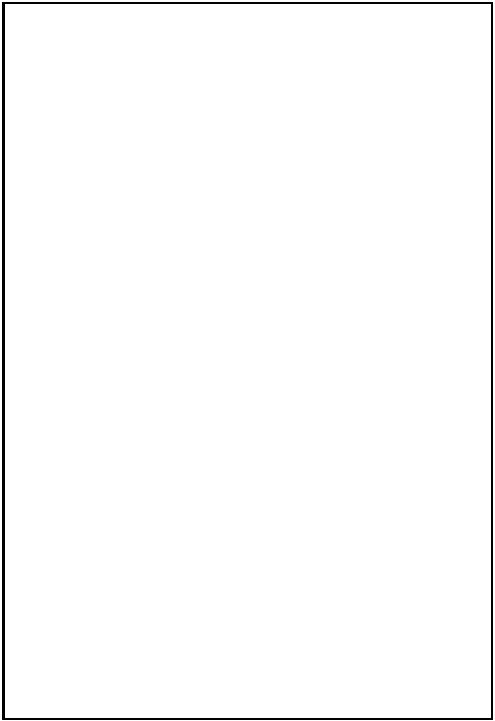

\vspace{9pt}
\framebox[65mm]{\rule[-21mm]{0mm}{93mm}}
\caption{Difference of $S({\bf q},\omega)$ (upper panel) and
$\chi''({\bf q},\omega)$ (lower panel) at ${\bf q}=(\pi,\pi)$
above and below $\rm T_c = 52$K for $\rm YBa_2 Cu_3 O_{6.5}$.}
\end{figure}

\begin{figure}[htb]
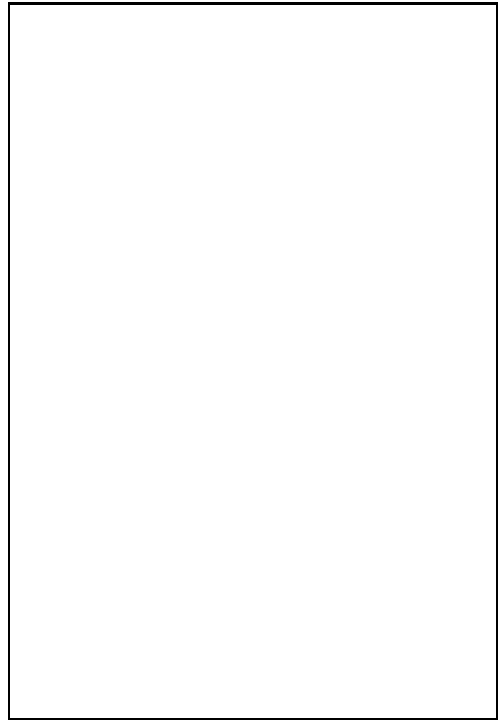

\framebox[65mm]{\rule[-21mm]{0mm}{93mm}}
\caption{Difference of $S({\bf q},\omega)$ (upper panel) and
$\chi''({\bf q},\omega)$ (lower panel) at ${\bf q}=(\pi,\pi)$
above and below $\rm T_c = 67$K for $\rm YBa_2 Cu_3 O_{6.7}$.}
\end{figure}

Our primary interest was the influence of superconductivity on the 
spin excitation spectrum. In Figs. 1 and 2 we have therefore plotted 
the difference of the intensity measured at low temperatures below 
$\rm T_c$ and the normal-state intensity above $\rm T_c$ for two 
$\rm YBa_2 Cu_3 O_{6+x}$ crystals of different oxygen 
concentrations, $\rm x \sim 0.5$ ($\rm T_c$=52K) and $\rm x 
\sim 0.7$ ($\rm T_c$=67K). The data were taken at the in-plane 
wavevector ${\bf q} = (\pi,\pi)$, and at the out-of-plane 
wavevector corresponding to the maximum of the sinusoidal 
magnetic structure factor. (For further discussions of the structure 
factor, see, {\it e.g.}, Ref. \cite{fong96}.) A subset of these data has been 
reported in Ref. \cite{fong97}, in arbitrary units. The data of Fig. 2 are 
also consistent with 
those of Ref. \cite{dai96}. Here we have plotted the data in two different 
ways, both as the scattering function $\rm S({\bf q},\omega)$ 
which is directly proportional to the magnetic cross section, and as 
the imaginary part of the dynamical susceptibility which is related to 
$\rm S({\bf q}, \omega)$ through the fluctuation-dissipation 
theorem: $\chi''({\bf q},\omega) = [1-exp(-\hbar\omega/k_BT)] 
S({\bf q}, \omega)$. (The unit conventions are the same as in Ref.
\cite{hayden96} and differ by $3\pi\mu_B^2/2$ from those of Ref. 
\cite{fong96}.)

\begin{figure}[htb]
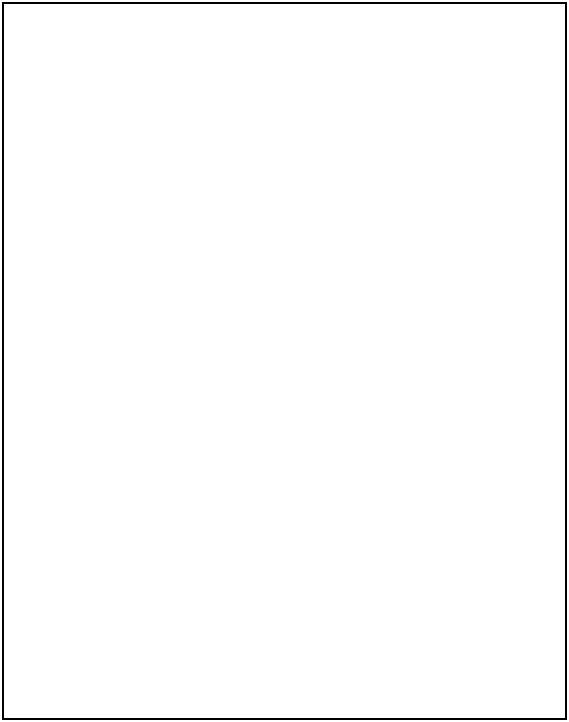

\vspace{9pt}
\framebox[75mm]{\rule[-21mm]{0mm}{93mm}}
\caption{Temperature dependence of $\chi''({\bf q},\omega)$ at 
$\hbar \omega=25$ meV
for $\rm YBa_2 Cu_3 O_{6.5}$ (upper panel) and at $\hbar \omega=33$ meV
for $\rm YBa_2 Cu_3 O_{6.7}$ (lower panel). The closed (open) symbols
represent data taken with a polarized (unpolarized) beam.}
\end{figure}

\begin{figure}[htb]
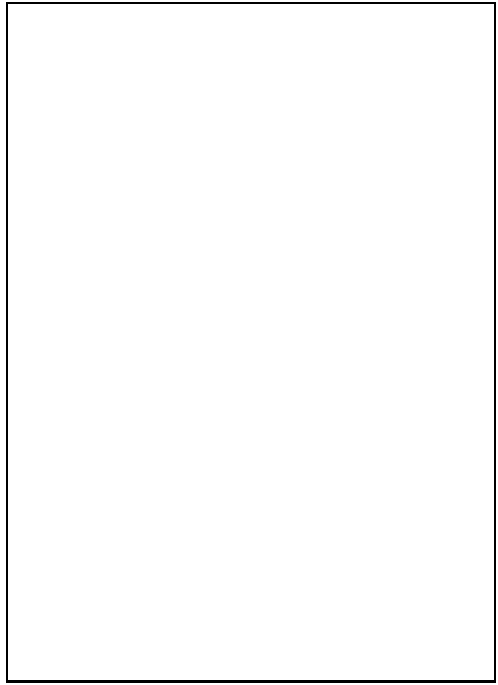

\vspace{9pt}
\framebox[65mm]{\rule[-21mm]{0mm}{88mm}}
\caption{$\rm T_c$-dependence of the enhanced part of the dynamical
susceptibility, $\int d\omega \chi''_+({\bf q},\omega)$, and the resonance 
energy $\rm E_{\rm res}$.}
\end{figure}

Clearly, the dynamical susceptibility below $\rm T_c$ exceeds
that above $\rm T_c$ over a certain energy range, centered around 
25 meV for $\rm YBa_2 Cu_3 O_{6.5}$ and 33 meV for $\rm YBa_2 Cu_3 O_{6.7}$.  
Fig. 3 shows that for both doping levels the enhancement of the susceptibility
is strongly correlated to the onset of superconductivity.
For $\rm YBa_2 Cu_3 O_{6.7}$ the energy range over which this 
enhancement is observed is limited only by the experimental resolution. 
The intrinsic linewidth of the enhanced part of the dynamical susceptibility, 
$\chi_+''({\bf q},\omega)$, is therefore indistinguishable from zero, as 
it is for $\rm YBa_2 Cu_3 O_7$ where $\chi_+''({\bf q},\omega)$ is 
centered around 40 meV. By contrast, $\chi_+''({\bf q},\omega)$ is somewhat 
broadened in $\rm YBa_2 Cu_3 O_{6.5}$.

\begin{figure}[htb]
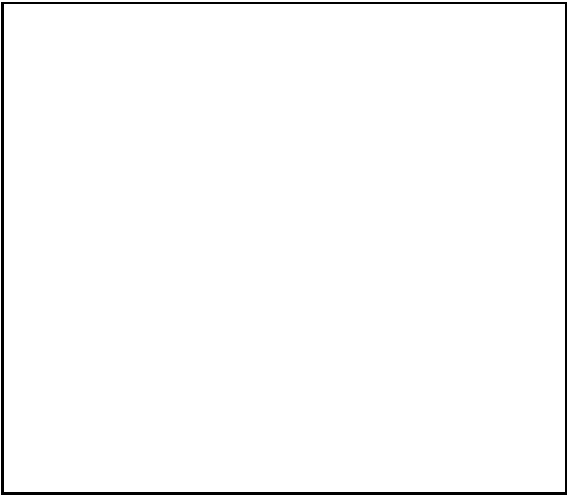

\vspace{9pt}
\framebox[75mm]{\rule[-21mm]{0mm}{63mm}}
\caption{Energy integral of the difference of $S({\bf q},\omega)$ 
above and below $\rm T_c$ at ${\bf q}=(\pi,\pi)$ and 
$0 \leq \hbar \omega \leq 50$ meV, for different doping levels.}
\end{figure}

In Fig. 4 we have summarized our observations. The integrated spectral
weight of the enhanced part of the susceptibility, 
$\int d\omega \chi_+''({\bf q},\omega)$,
decreases as a function of increasing $\rm T_c$, whereas $\rm E_{\rm res}$, 
the energy around which $\chi_+''({\bf q},\omega)$ is peaked, 
increases. While Fig. 3 shows that the ratio of the resonance spectral weight 
to the spectral weight of the normal-state spin excitations increases 
strongly with increasing doping, it is interesting to 
note that the resonance spectral weight actually decreases on an absolute 
scale with increasing carrier concentration. The functional dependence of 
$\rm E_{\rm res}$ on $\rm T_c$ (or doping, which depends monotonically 
on $\rm T_c$) is obviously not well defined by the three data points, 
but the qualitative trend contrasts sharply with the weak doping dependence 
of the energy gap directly determined by photoemission spectroscopy 
\cite{harris96}. This discrepancy was predicted in the model of Zhang and 
collaborators \cite{zhang95} where the resonance energy is not tied 
to the gap but is directly related to the doping level. It does not rule out
the gap model, however. Millis and Monien \cite{millis96} have shown that
the resonance energy can be lower than the gap energy due to final state 
interactions of the quasiparticle-quasihole pair. These interactions are
expected to increase as the doping level is reduced. A more clearcut
picture may emerge when the present data are compared in detail to 
quantitative predictions of the resonance energy and integrated spectral 
weight \cite{millis96,zhang95}.

Finally, we turn to the relation between the normal-state spin excitations and
the resonance in the superconducting state. We are motivated by the total moment
sum rule which requires that the
the integral $\int d {\bf q} d \omega S({\bf q},\omega)$ is weakly temperature
dependent (and temperature independent for local-spin models).
The difference of $S({\bf q},\omega)$ in the normal
and superonducting states is plotted in the upper panels of Figs. 1 and 2
for ${\bf q} = (\pi,\pi)$. For $\rm YBa_2 Cu_3 O_{6.5}$ the positive
and negative areas are equal to within our resolution, that is, the total
moment sum rule is exhausted for ${\bf q} = (\pi,\pi)$ and in the energy
range investigated in our experiment ($0 \leq \hbar \omega \leq 50$ meV).
By contrast, in both $\rm YBa_2 Cu_3 O_{6.7}$ and $\rm YBa_2 Cu_3 O_7$
the positive part of the difference clearly exceeds the negative
part in this domain of $({\bf q},\omega)$. This situation is summarized
in Fig. 5. For the total moment sum rule
to be satisfied, the intensity of the resonance mode must be drawn from a
much wider range of $({\bf q},\omega)$. Further investigations, especially
at higher energies, are clearly warranted.

\end{document}